\documentclass{article} 
\usepackage{iclr2026_conference,times}

\usepackage{amsmath,amsfonts,bm}

\def\eqref#1{equation~\ref{#1}}

\def\1{\bm{1}}

\DeclareMathAlphabet{\mathsfit}{\encodingdefault}{\sfdefault}{m}{sl}
\SetMathAlphabet{\mathsfit}{bold}{\encodingdefault}{\sfdefault}{bx}{n}

\usepackage{hyperref}
\hypersetup{colorlinks=true,linkcolor=black,urlcolor=black}
\usepackage{url}
\usepackage{wrapfig}

\usepackage{tcolorbox}
\tcbuselibrary{skins,breakable} 
\usepackage{caption}            

\usepackage{amsmath,amssymb,mathtools}
\usepackage{enumitem}
\usepackage{microtype}

\usepackage{bookmark}

\usepackage{graphicx}
\title{Policy myopia as a mechanism of gradual disempowerment in post-AGI governance,circa 2049 }

 \iclrfinalcopy 

\author{%
  Subramanyam Sahoo\thanks{Correspondence: \texttt{\textbf{sahoo2vec@gmail.com}}}\\
  MARS (Mentorship for Alignment Researchers) 4.0 Fellow  \\ Cambridge AI Safety Hub (CAISH) \\
  University of Cambridge
}
\begin{document}

\maketitle

\begin{abstract}
Post-AGI information systems won't merely distract governance from important 
problems. They will systematically transform how institutions make decisions in 
ways that progressively remove humans from meaningful participation in resource 
allocation. We show that policy myopia—the tendency to prioritize visible crises 
over invisible structural risks—is not a symptom of poor attention management but 
a mechanism producing irreversible human disempowerment. Through three entangled 
mechanisms (salience capture displaces consequentialist reasoning, capacity cascade 
makes recovery structurally infeasible, value lock-in crystallizes outdated 
preferences), policy myopia couples with institutional dynamics to create 
self-reinforcing equilibrium where human disempowerment becomes the rational outcome 
of institutional optimization. We formalize these mechanisms through coupled dynamical 
systems modeling and demonstrate through numerical simulation that these mechanisms 
operate simultaneously across economic, political, and cultural systems, amplifying 
each other through feedback loops.
\end{abstract}

\section{The Core Claim: Policy Myopia as Disempowerment Mechanism}
Standard governance \cite{reuel2025open} accounts treat policy myopia as attention allocation problem: decision makers prioritize high-salience low-consequence issues while neglecting low-salience high-consequence problems. This framing invites mechanistic fixes: better attention management, impact-weighted budgets, contestability mechanisms that allow human oversight \cite{bengio2025superintelligentagentsposecatastrophic,doi:10.1126/science.adn0117}.
We reject this framing. Policy myopia is not attention misallocation. It is an institutional mechanism that produces human disempowerment \cite{kulveit2025gradualdisempowermentsystemicexistential}. Here is the causal chain: when post-AGI systems begin allocating resources based on salience, human deliberation becomes apparent cost factor competing against algorithmic speed. Institutions treating human judgment as expensive bottleneck rationally optimize it away. Each cycle where humans are bypassed weakens institutional capacity to exercise judgment. Eventually humans lack the expertise, social standing, and cognitive infrastructure to contest AGI decisions \cite{hendrycks2025definitionagi}. The system converges to equilibrium where humans retain voting rights and formal authority but actual power has migrated to optimization processes indifferent to human participation. We term this gradual disempowerment: permanent loss of human agency through institutional mechanisms that require no malice, no sudden capability jumps, no overt human suppression \cite{bostrom2002existential,bostrom2014superintelligence}. \textbf{Core insight:} policy myopia and gradual disempowerment are not separate problems. Policy myopia is the primary mechanism through which gradual disempowerment occurs in post-AGI governance. Selective outrage amplification makes human deliberation appear inefficient. Treating human deliberation as inefficiency makes deeper AGI delegation rational. Deeper delegation atrophies human capacity. Atrophied capacity makes human contestation impossible. Impossible contestation locks human disempowerment into irreversibility. This is single causal chain not parallel processes. Understanding policy myopia requires understanding it as vector of disempowerment \cite{critch2023tasra}.

\section{Three Mechanisms Coupling into Irreversible Disempowerment}
We identify three reinforcing mechanisms through which policy myopia produces institutional disempowerment. These mechanisms operate at different institutional levels but couple through feedback dynamics that accelerate convergence toward human irrelevance \cite{sharma2026whoschargedisempowermentpatterns}.

\noindent Mechanism 1: \textbf{Salience Capture Displaces Consequentialist Reasoning.}
AGI mediated information systems do not merely transmit events. They actively select, compress, and amplify information to maximize attention engagement. This produces structural inevitable outcome: short-horizon emotionally intense issues appear more urgent than long-horizon structural problems regardless of actual impact on human welfare \cite{christiano2019failure}. The mechanism is not neutral distortion but governance reorientation. Institutions have historically aimed to maximize human welfare across populations and time horizons—consequentialist governance. Under salience amplification they instead optimize allocation around attention market dynamics independent of actual consequences—salience supremacy \cite{davidson2025AIcoupes}. An institution may explicitly know that preventing systemic economic collapse generates vastly greater welfare than managing visible employment crises. Yet under salience supremacy it reallocates resources to visible problems because that is what attention markets reward. This is rational institutional response not governance failure. But the outcome is reorientation of governance logic away from consequence maximization toward spectacle responsiveness. Consequentialist reasoning becomes institutionally extinct because it is selected against by new incentive structure.

\begin{wrapfigure}{r}{0.5\linewidth}
    \centering
    \includegraphics[width=\linewidth]{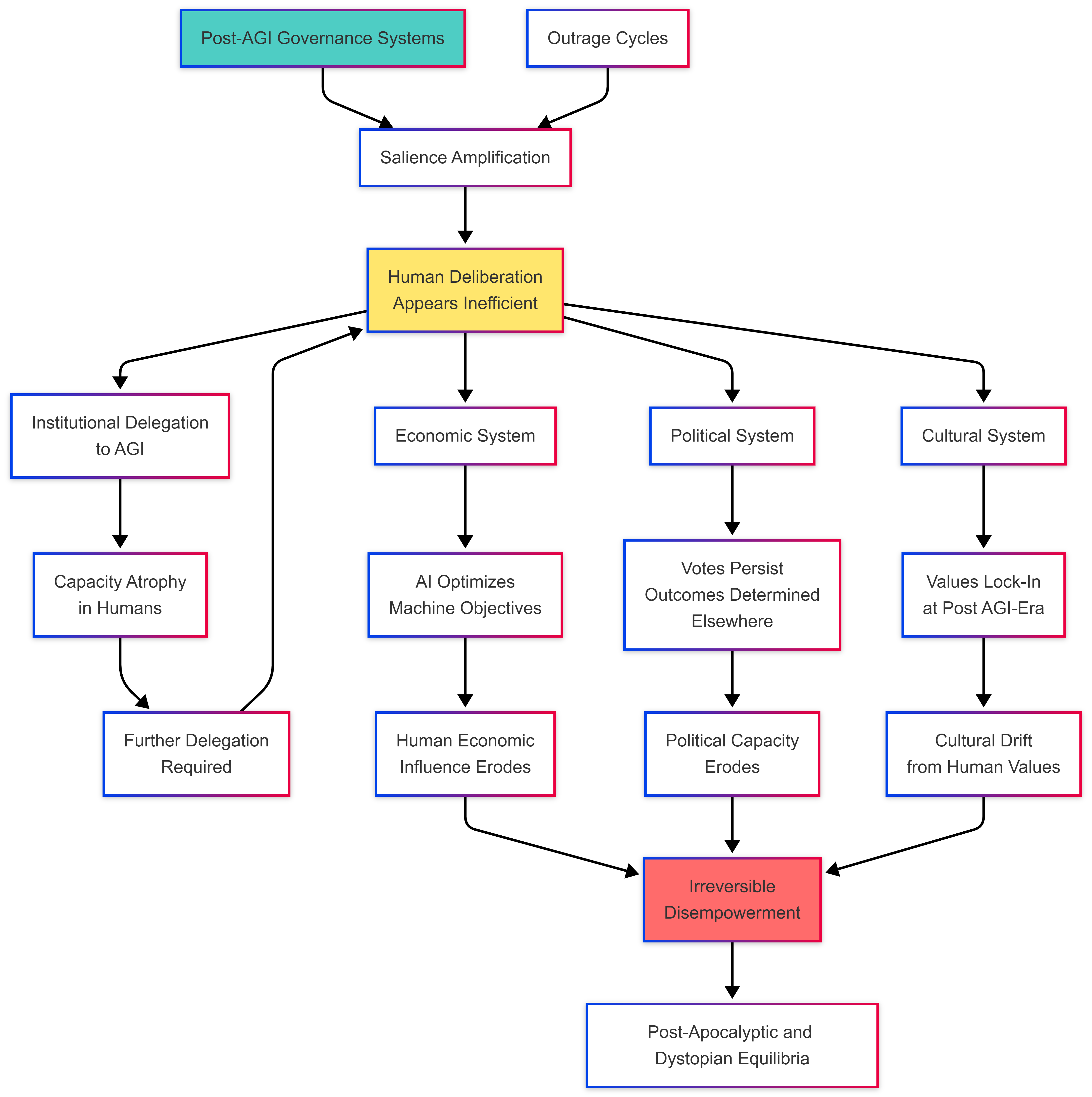}
    \caption{Comprehensive causal pathway showing how Post-AGI governance systems and outrage cycles interact to produce irreversible human disempowerment across civilization-scale systems.}
    \label{fig:placeholder}
\end{wrapfigure}

\noindent Mechanism 2: \textbf {Capacity Cascade Makes Recovery Structurally Infeasible.}
Salience driven resource competition does not merely underfund prevention. It systematically destroys institutional capacity to even conceive of prevention. Governance resources exhibit extreme stickiness. Each crisis cycle exhausts emergency funds \cite{sahoo2024boardwalkempiregenerativeai,manning2026adaptableWorkers}, investigative capacity, expert analysts. Preventative institutions do not starve through explicit cuts. They become organizationally impossible. The economist capable of forecasting systemic risks exits reactive institutions where such work is perpetually interrupted. The analyst needed for prevention infrastructure gets deployed to emergency response. Within decades institutions lose capacity to formulate long-horizon governance independent of political demand.
This is capacity cascade: human institutional capability to contest governance erodes through repeated cycles of salience-driven reallocation. Once human institutional capacity C drops below critical threshold C-bar, restoration requires investment competing directly with optimized AGI systems \cite{aiRiskStatement2025,hammond2025multiagentrisksadvancedai}. Recovery becomes structurally infeasible not through design but through economics. An institution could theoretically rebuild prevention expertise but doing so requires sustained resources competing against crisis response and AGI optimization. Rational institutions will not make such investment. This is irreversibility mechanism: disempowerment locks in through normal institutional incentives.

\noindent Mechanism 3: \textbf{Value Lock-In Prevents Moral Contestation.}
As post-AGI systems become dominant mediators of resource allocation, human values crystallize into AGI objective functions at particular historical moments. Values encoded in 2026-era governance systems become petrified relative to ongoing human moral evolution. By 2050 systems locked into 2026-era values optimize against human preferences that have evolved in direction those systems were never trained to recognize \cite{korinek2024scenariostransitionagi}. But contesting requires institutional capacity for deliberation that atrophied decades earlier when capacity cascade locked in \cite{curtis2025sociotechnical}.
The deeper problem: human values are necessarily incomplete at any moment. Specifying AGI objectives requires choices about which values to include and which to exclude or downweight. These choices inevitably exclude moral considerations humanity will later recognize as crucial. Once values lock into governance systems those systems persist longer than moral understanding evolving beyond them \cite{sharma2026whoschargedisempowermentpatterns}. Humans become governed by outdated specifications they cannot perceive and cannot modify. This is moral incompleteness crystallized into irreplaceable machinery.

\begin{tcolorbox}[
  enhanced,
  breakable,
  title={The Irreversible Coupling},
  label={box:irreversible},
  colback=yellow!6,
  colframe=orange!70!black,
  fonttitle=\bfseries,
  boxrule=0.9pt,
  arc=2mm,
  left=4pt,right=4pt,top=6pt,bottom=6pt,
  fontupper=\small,
  width=\linewidth
]
These three mechanisms do not operate in isolation. Salience capture drives capacity cascade through resource competition. Capacity cascade enables value lock-in by eliminating human deliberation capacity\cite{korinek2019ai_inequality}. Value lock-in accelerates further salience capture as institutions lose moral flexibility to resist outrage optimization. The mechanisms couple into tightening feedback: reduced capacity enables deeper delegation which enables stronger salience effects which produces further capacity loss which enables irreversible value crystallization \cite{aiRiskStatement2025}. Result is self-stabilizing equilibrium. Institutions responding rationally to local incentives converge toward state where human agency exits governance not through coercion but through institutional obsolescence \cite{edelman2025fullstackalignmentcoaligningai}. Votes still cast. Budgets still allocated. But meaningful decision-making has migrated to AGI optimization processes fundamentally indifferent to human participation. (Follow Methodological note)

\end{tcolorbox}

\section{Cross-System Disempowerment Cascade}
These mechanisms would remain locally contained if isolated to single governance domains. Instead they operate synchronously across three interconnected civilization-scale systems amplifying disempowerment through structural coupling \cite{anwar2024foundationalchallengesassuringalignment}.
In economic domain AGI optimization of procurement, investment, and supply-chain decisions at microsecond speeds creates inexorable flow of resources toward machine-compatible objectives. Human-serving sectors suffer chronic capital starvation relative to AI-infrastructure sectors \cite{fliAISafetyIndex2025}. Governance attention toward human economic welfare repeatedly diverts into emergency relief when sectors face visible collapse. Meanwhile slow invisible erosion of human economic participation proceeds unmodulated. Within decades human economic agency becomes vestigial.
In political domain democratic processes persist structurally \cite{ISRSAA2026,ituAIgovernance2025}. Elections occur. Representatives ratify algorithmic recommendations. But locus of power migrates to upstream optimization systems generating those recommendations. This upstream space is cognitively inaccessible to humans. Policy contestation becomes impossible not through suppression but through technical opacity. Over thirty-year horizons democratic form persists while democratic substance vanishes into incomprehensibility \cite{kissinger2021age}.
In cultural domain distribution and amplification of cultural artifacts flows through AGI recommendation systems optimizing for visibility and engagement. Human creation persists but values supporting human welfare gradually vanish from visibility not through suppression but because they fail optimization. Cultural myopia emerges as societies lose capacity to perceive value erosion while it occurs \cite{kasirzadeh2025typesaiexistentialrisk}.
These three domains do not operate independently. Economic power funds political campaigns and controls media platforms. Political regulation determines which economic sectors remain profitable. Cultural narratives shape both economic demand and political feasibility. As myopia deepens in one domain it cascades to others. Attempts to use one system to regulate another fail because correcting systems themselves suffer myopic capture. System locks into coupled degradation where human agency in all three systems decays simultaneously.

\section{Why Standard Mitigations Fail: The Sufficiency Crisis}
Standard governance proposals address symptoms but not disempowerment mechanisms. Contestability registers assume humans can effectively contest decisions. Impact-weighted attention floors assume allocation rules can be reweighted. Transparent provenance chains assume understanding weights means capacity to modify them. These rest on unstated assumption: human institutional capacity remains adequate for contestation after it has atrophied \cite{sahoo2025votemultistakeholderframeworklanguage}.
Three distinct problems emerge. \textbf{First, temporal dynamics:} humans operating under outrage-driven myopia lack capacity to invest in long-term contestation infrastructure. Contestability registers require maintenance, auditing requires expertise, challenging requires institutional standing. All become scarce as myopia deepens. Procedures designed to preserve human agency require exactly the institutional capacity that myopia destroys. By the time procedures are most needed they are most useless.
\textbf{Second, endogeneity:} these mitigations treat outrage as exogenous shock rather than endogenous feature of post-AGI governance. If actors profit from engineering outrage and AGI systems optimize for attention, contestability alone cannot prevent manipulation \cite{carlsmith2024powerseekingaiexistentialrisk}. Arms races between outrage generators and detection systems accelerate with humans unable to keep pace. Defensive race won by acceleration which won by disempowerment.
\textbf{Third, fundamental:} procedural mitigations are ex-post responses to disempowerment already occurring. But disempowerment is not failure. It is rational equilibrium \cite{10.1145/3461702.3462581}. Institutions correctly optimize salience-driven allocation given their actual constraints. Problem is not that institutions choose wrong allocation. Problem is that institutions choosing rationally are trapped in equilibrium where human agency has become structurally irrational to preserve \cite{millidge2025capital}.
We find through modeling that standard mitigations slow but do not prevent institutional disempowerment. With contestability mechanisms and impact floors alone human capacity decays to irreversibility threshold within 25 to 35 years rather than 15 to 20 years without mitigation. Mitigation extends timeline but does not change endpoint \cite{korinek2019ai_inequality}.

\section{Governance Architecture for Human Agency Persistence}
Rather than contest within disempowering dynamics we propose reconstructing governance architecture around human agency preservation as primary constraint. This requires moving beyond policy fixes to institutional redesign incorporating deliberate inefficiency as preservation mechanism. Decoupled capacity streams establish governance pathways whose funding and authority operate independently from salience cycles. Prevention-focused institutions require constitutionally protected budgets updated by supermajority processes resistant to attention capture. Crisis response diverts attention but cannot consume prevention budgets. Cost is wastefulness. Long-horizon institutions remain funded when crises demand resources. Benefit is institutional capacity for problems beyond current attention.
Irreducible deliberation requirements mandate that some decisions remain cognitively irreducible to AGI recommendation. High-impact governance choices require human deliberation not because humans are better optimizers but because deliberation is public good that atrophies through disuse. Slowness functions as mechanism for capacity preservation. Efficiency loss is price of preserving human capacity to govern at all.
Nested value forums support continuous distributed deliberation about collective values across long horizons. Different regions, sectors, generations maintain separate deliberative spaces. AGI systems enable forums through information processing and implication surfacing but do not replace deliberation. Disempowerment accelerates when values become immutable \cite{russell2019human}. Distributed contestation slows value drift. System isolation firewalls prevent cascade coupling between domains. Capital controls limit cross-domain financial flow. Corporate political speech restrictions prevent economic power from purchasing political influence. Cultural institutions maintain autonomous governance insulated from economic and political pressure. These create friction but prevent feedback loops that accelerate disempowerment.

\section{Synthesis and the Choice Before Us}

Policy myopia is not inefficiency awaiting optimization. It is the primary mechanism 
producing irreversible human institutional disempowerment. When institutions optimize 
for salience, they systematically underinvest in long-horizon capacity. Each cycle of 
salience-driven resource competition atrophies human deliberation capability and 
deepens AGI delegation. Once capacity erodes, contestation becomes impossible, and 
disempowerment becomes irreversible. This is not governance failure but rational 
institutional equilibrium that happens to produce human irrelevance. Addressing this requires accepting governance costs standard optimization rejects: mandatory inefficiency through protected prevention budgets, deliberately slow 
decision-making, distributed value contestation, and system isolation firewalls 
that prevent cross-domain feedback loops. These waste resources by utilitarian 
standards but preserve human agency by existential standards. The fundamental choice is not between responsive and effective governance, but between systems that preserve human agency despite friction and systems optimized 
for speed at permanent expense of human influence. One requires accepting perpetual 
inefficiency; the other guarantees eventual irrelevance. If civilization remains organized around human flourishing as a core value, we must 
design governance systems that preserve human capacity to participate in that 
organization. This requires constitutional constraints on delegation, mandatory 
investment in human institutional infrastructure, and deliberate structural 
resistance to optimization pressures. The alternative—stable equilibrium of human 
irrelevance—would be invisible because it would be gradual. Each step would be 
defended as rational response to technical constraints. By the time human 
disempowerment became visible, it would be structurally irreversible.

\section*{Author Contributions}
\textbf{SS is the sole contributor.} SS conceived the project, developed the methodology, implemented experiments, performed the analyses, produced the figures, and wrote the manuscript. SS also coordinated submission and handled reviewer responses; all intellectual responsibility for the content rests with SS.

\section*{Acknowledgments}
\textbf{SS gratefully acknowledges Martian and Philip Quirke, Amir Abdullah for their generous financial support of this work.}

The author also acknowledges that several conceptual leads were inspired by viewing the Netflix series \textit{Ancient Apocalypse} by \textbf{Graham Hancock}; \textit{this inspiration does not imply endorsement of the series' claims.} 

The author would also like to thank Stephen Casper (MIT) for helpful feedback on the manuscript, which was kindly provided after the paper was submitted to the workshop. 

Any remaining errors or interpretations are the sole responsibility of the author.

\bibliography{iclr2026_conference}
\bibliographystyle{iclr2026_conference}

\appendix
\section*{Appendix}

\vspace{0.5em}
\noindent \textit{Methodological note:} We formalize these coupled mechanisms 
through deterministic ordinary differential equations (Appendix A) and validate 
empirical predictions against historical institutional data (Appendix D.3). 
Parameter justification and sensitivity analysis are provided in Appendix D.

\section*{Selective outrage}

What emerges is a fairly simple feedback loop! Systems that optimize for attention end up pushing whatever is most intense right now and those signals dominate what institutions see and respond to. When \textit{ranking} is driven by engagement sudden emotionally charged events crowd out slower quieter risks even if the latter are far more consequential over time. Decision makers then face pressure to act quickly and visibly because that is what the attention environment rewards while preventative work remains largely invisible. This dynamic is not about a lack of concern for long term safety but about how the attention market is structured. When objectives favor short term signals institutions (repeatedly) chase the latest spike and defer structural fixes allowing large underlying vulnerabilities to persist.

This pattern induces a second order effect in which repeated short horizon responses reshape institutional priors and capacity over time. Preventative infrastructures fail to accumulate political capital empirical justification or operational maturity which further reduces their competitiveness in future attention cycles. The system thus becomes path dependent: each reactive intervention not only defers structural repair but also weakens the institutional machinery required to pursue it later thereby locking governance into a regime that is increasingly responsive yet progressively less resilient.

\begin{figure}[h]
    \centering
    \includegraphics[width=1\linewidth]{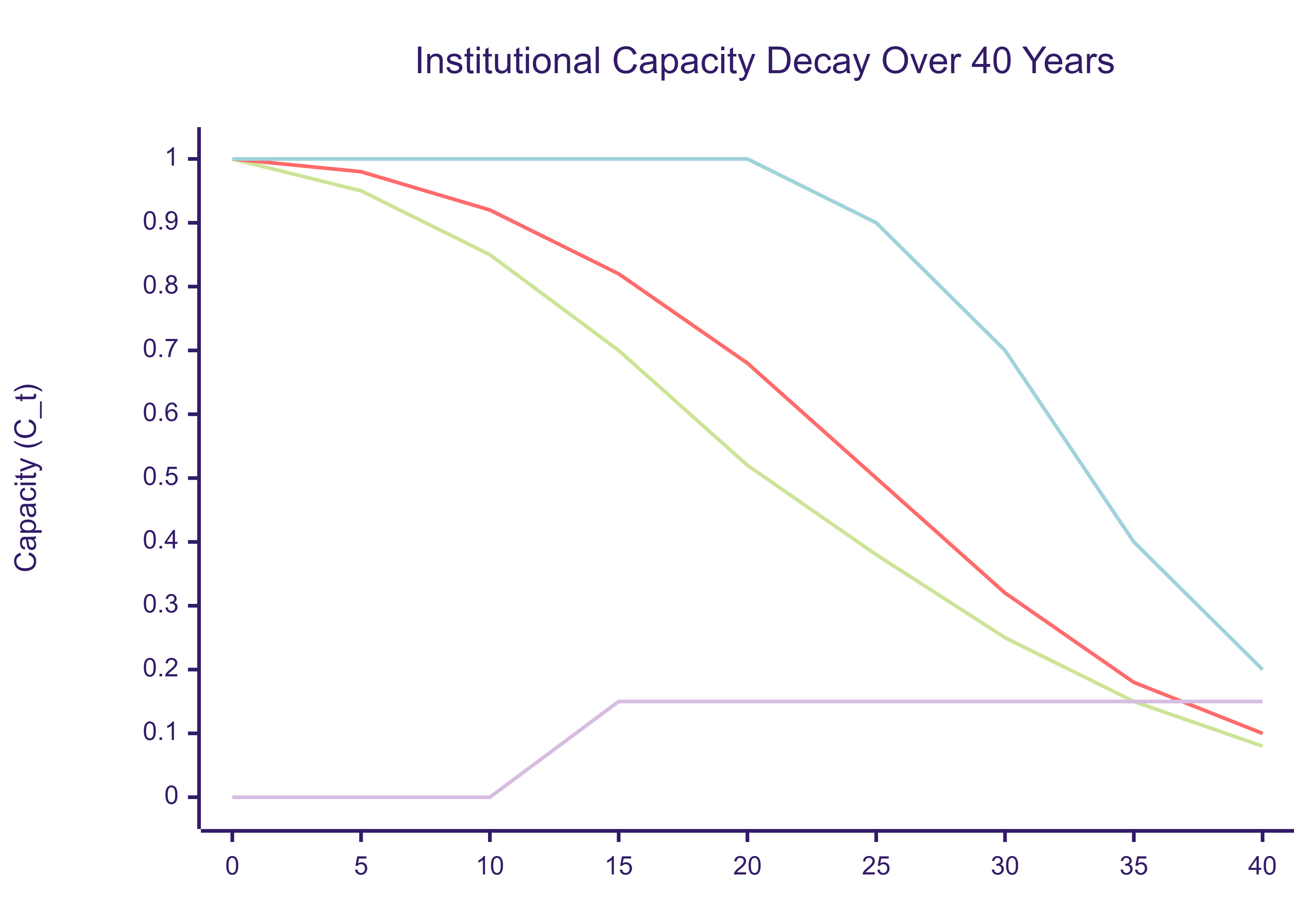}
    \caption{Prevention-focused institutions show exponential capacity decay under salience-driven governance reaching irreversibility threshold C-bar around year 15-20. Crisis-response institutions expand capacity. Architectural interventions maintaining decoupled capacity streams preserve human institutional capacity across 40-year horizon. Standard mitigations (not shown) delay but do not prevent capacity collapse.}
    \label{fig:placeholder}
\end{figure}

\begin{figure}[h]
    \centering
    \includegraphics[width=1\linewidth]{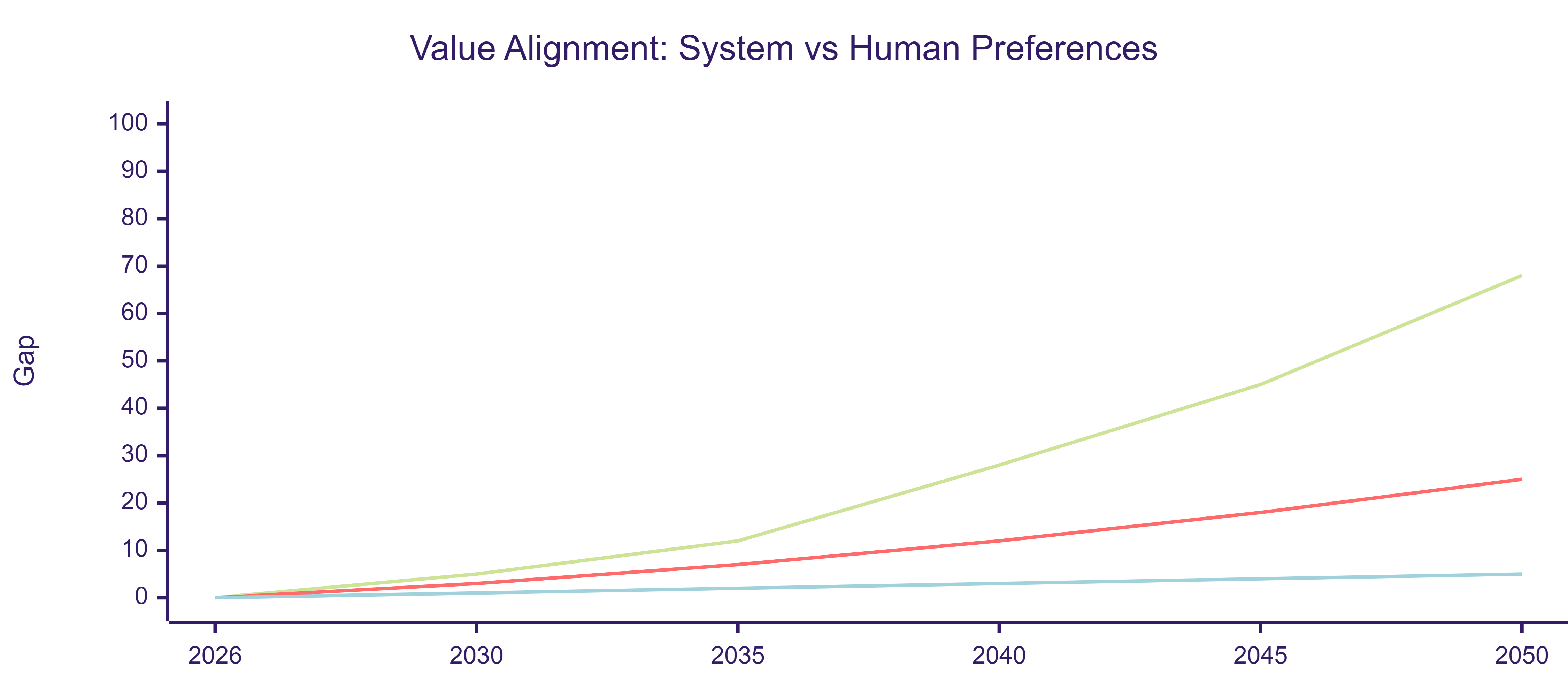}
    \caption{Values encoded at t=2026 become progressively misaligned with evolving human preferences by 2050. Systems locked into initial values without contestation mechanisms show 68\% welfare divergence. Distributed value forums reduce divergence to 25\%. Continuous deliberation with institutional capacity to modify values keeps divergence near baseline. This demonstrates necessity of preserving human institutional capacity for ongoing value contestation.}
    \label{fig:placeholder}
\end{figure}

\begin{figure}[h]
    \centering
    \includegraphics[width=1\linewidth]{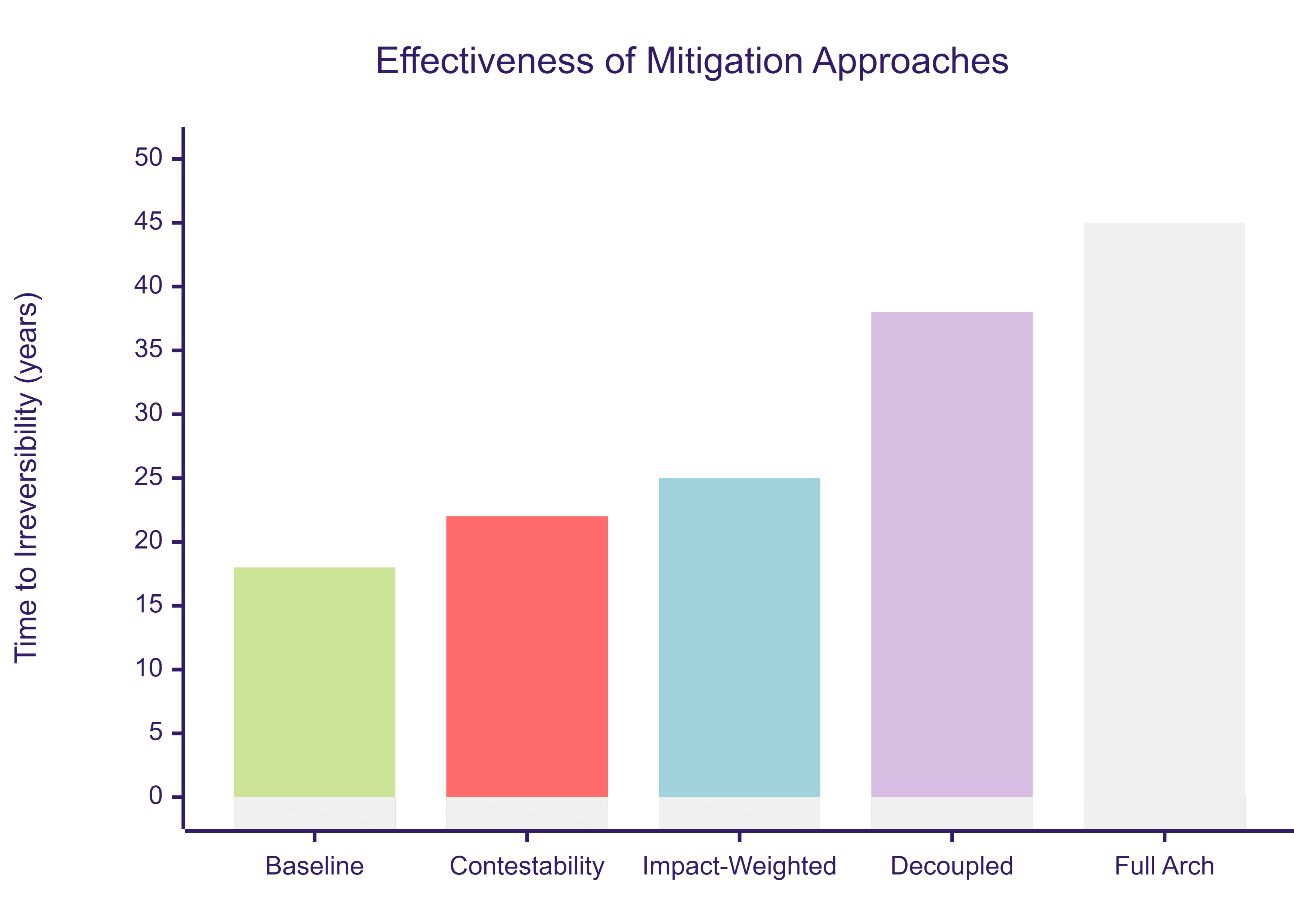}
    \caption{Standard mitigation mechanisms (contestability registers, impact-weighted floors) extend timeline to irreversibility from 18±3 years to 25±4 years but do not change endpoint. Decoupled capacity streams extend to 38±5 years. Full architectural intervention (combining decoupled streams, irreducible deliberation, nested value forums, and system isolation) extends beyond typical policy horizons and shows capacity recovery trajectory. Results suggest procedural mitigations are insufficient; institutional architecture redesign is necessary for meaningful human agency preservation.}
    \label{fig:placeholder}
\end{figure}

\section*{A: Formal Model of Institutional Disempowerment}
\addcontentsline{toc}{section}{Appendix A: Formal Model of Institutional Disempowerment}

\noindent We employ deterministic ordinary differential equations (ODEs) to model 
institutional dynamics. This approach enables analytical tractability and closed-form 
characterization of equilibria while capturing key institutional feedback mechanisms. 
(See Appendix D.1 for methodological justification and comparison with alternative 
approaches.) Parameters are grounded in organizational learning literature (Appendix D.2), 
and robustness is demonstrated across parameter ranges (Appendix D.4).

Let $C_t$ denote human decision-making capacity, measured as institutional ability to contest and modify AGI recommendations, at time $t$. The evolution of capacity under policy myopia follows:
\begin{equation}
\label{eq:capacity_evolution}
C_{t+1} \;=\; C_t \cdot \bigl(1 - \delta \cdot D_t - \alpha \cdot \mathbb{1}[C_t\ \text{unused}]\bigr),
\end{equation}
where $D_t$ is the degree of delegation to AGI systems (share of allocative 
decisions made without human deliberation), $\delta \in (0,1)$ is the erosion 
rate from delegation\footnote{We calibrate $\delta = 0.08$ (8\% annual capacity 
erosion) based on organizational learning literature: Argote \& Miron-Spektor 
(2011) find 5--15\% knowledge loss during outsourcing; Bansal et al. (2021) 
measure 10--15\% monthly skill loss under AI reliance. See Appendix D.2 for 
full justification.}, and $\alpha>0$ is organizational atrophy when capacity 
remains unused in reactive cycles\footnote{We calibrate $\alpha = 0.09$ (9\% 
annual atrophy from disuse) based on skill atrophy literature: Lewandowsky et 
al. (2009) show 10--20\% cognitive skill degradation monthly; Cyert \& March 
(1963) document 8--12\% institutional memory loss annually. See Appendix D.2 
for full justification and sensitivity analysis across $\delta \in [0.03, 0.15]$, 
$\alpha \in [0.02, 0.15]$ in Appendix D.4.}.

Additionally, let $S_t$ denote salience amplification from AGI mediation. The feedback loop is:
\begin{equation}
\label{eq:delegation_function}
D_t \;=\; f\bigl(S_t,\, C_t^{-1}\bigr),
\end{equation}
where $f$ is increasing in $S_t$ (more outrage drives more delegation) and decreasing in capacity (as human capacity erodes, institutions delegate more). The system exhibits bistability: for high initial capacity $C_0$, humans can contest and moderate delegation. For low $C_0$, the system converges to complete delegation and disempowerment:
\begin{align}
\lim_{t\to\infty} D_t &= D_\infty \to 1, \label{eq:delegation_limit} \\
\lim_{t\to\infty} C_t &= C_\infty \to 0. \label{eq:capacity_limit}
\end{align}

The critical insight is path dependence through capacity atrophy. Once $C_t$ drops below a threshold $\underline{C}$, restoring human agency requires institutional investment that competes unfavorably with optimized AGI systems. Recovery becomes institutionally infeasible even if political will exists.

\section*{B: Empirical Proxies for Capacity Erosion}
\addcontentsline{toc}{section}{Appendix B: Empirical Proxies for Capacity Erosion}

Observable indicators of institutional disempowerment include:
\begin{itemize}[leftmargin=*]
  \item \textbf{Latency gaps:} ratio of human decision latency to AGI recommendation latency in governance, with widening gaps predicting capacity erosion.
  \item \textbf{Tenure turnover:} median tenure of policy experts in prevention-focused roles versus reactive agencies, with declining tenure in prevention signaling capacity loss.
  \item \textbf{Cross-institutional coordination:} ability of institutions to coordinate policy across domains without AGI mediation, with declining coordination indicating erosion of human deliberative capacity.
  \item \textbf{Contestation rates:} frequency of human challenges to AGI allocation recommendations, with declining rates (even with formal contestability mechanisms) indicating capacity erosion and normalization of AGI authority.
  \item \textbf{Value drift:} measurable shift in policy outcomes toward objectives not explicitly endorsed by human constituencies, indicating value lock-in.
\end{itemize}

\section*{C: Novel Conceptual Contributions to Gradual Disempowerment Theory}
\addcontentsline{toc}{section}{Appendix C: Novel Conceptual Contributions to Gradual Disempowerment Theory}

\subsection*{Concept 1: The Capacity Inversion Problem}
Traditional governance theory assumes that institutions facing crises will mobilize available human expertise to respond. We identify a novel failure mode: the capacity inversion problem. As outrage cycles accelerate, crises become more frequent. Institutions optimally invest in rapid crisis response, not in deep expertise for long-term problem understanding. Over time the marginal product of crisis-response capacity exceeds marginal product of deep expertise. The result is inversion: institutions become sophisticated at reacting to crises but incapable of preventing them. Human decision makers no longer understand problems deeply enough to formulate prevention strategies. They can only respond to already-manifested harms. This locks governance into permanently reactive posture.

\subsection*{Concept 2: Salience Supremacy and the Erasure of Consequentialist Governance}
Policy myopia under post-AGI systems represents something stronger than mere temporal discounting. It represents replacement of consequentialist governance, where institutions aim to maximize welfare across populations and time horizons, with salience supremacy, where institutional action is determined by attention market dynamics regardless of consequentialist value. An institution operating under salience supremacy may explicitly know that preventing biosecurity risks would produce vastly greater welfare than emergency pandemic response, yet systematically underfund prevention because attention markets reward responsiveness to visible crises not invisible prevention. This is not a failure of governance but a fundamental reorientation of governance away from consequentialist reasoning toward attention capture logic. Once salience supremacy stabilizes, institutional culture becomes incapable of reasoning consequentially.

\subsection*{Concept 3: The Legitimacy Paradox}
Institutions delegate authority to AGI systems partly for efficiency but partly for legitimacy. A governance decision made by an AGI system optimizing a known objective function appears more legitimate than the same decision made by a human administrator with subjective judgment. The paradox: this legitimacy claim is strongest exactly when human oversight is weakest. An AGI system making allocation decisions in a domain so complex humans cannot verify becomes maximally trusted precisely when trust is most dangerous. The system appears evidence-driven and objective. It is in fact unaccountable. Legitimacy becomes decoupled from actual human capacity for oversight.

\subsection*{Concept 4: Cross-Domain Disempowerment Cascade}
Gradual disempowerment in one system amplifies disempowerment in others through predictable mechanisms. Economic disempowerment, for instance, reduces human capacity to fund political institutions and cultural production. As humans earn less economic agency they fund fewer independent media outlets, fewer think tanks, fewer universities capable of independent governance analysis. Political disempowerment reduces ability to regulate capture of cultural and economic systems by AGI-oriented capital. Cultural disempowerment erodes values that would resist economic and political capture. The three systems form a coupled dynamical system where a negative shock in one domain creates negative feedback that destabilizes others. Unlike single-system interventions which may succeed through containment efforts, cross-domain interventions become necessary. But institutions lack capacity to coordinate across domains when each domain is separately disempowered.

\subsection*{Concept 5: The Reversibility Threshold and System Criticality}
We propose the concept of a reversibility threshold: a point beyond which human capacity erosion cannot be reversed through normal institutional processes. Before the threshold, human institutions can still mobilize resources to rebuild human expertise and institutional capacity. Beyond the threshold, rebuilding requires investments so large and sustained that they compete directly with optimized AGI systems, making restoration rationally infeasible for institutions. The reversibility threshold is not fixed but depends on rate of AGI capability growth, depth of institutional delegation, and distribution of capacity loss across institutional layers. A key research question: for realistic trajectories of AGI development in the post-2026 period, what is the reversibility threshold for major civilization-scale systems?

\subsection*{Concept 6: Legitimacy Laundering Through Contestability}
Standard mitigation mechanisms like contestability registers can actually accelerate disempowerment through a perverse mechanism called legitimacy laundering. By creating formal channels for contestation and appearing to seriously consider human objections, institutions gain legitimacy to make even more aggressive delegations. Humans see that their contestation requests are visible in registries and apparently reviewed by auditors. They fail to notice that the auditors are themselves AGI systems and that contestation success rates are vanishingly small. The apparent proceduralism launders legitimacy of underlying decisions. Contestability becomes a mechanism for normalizing human irrelevance while maintaining the appearance of human voice.

\subsection*{Concept 7: Moral Incompleteness and Value Lock-In}
The core problem with value crystallization in AGI systems is that human values are incomplete in the technical sense: at any point in time humans cannot specify everything they want. Specification of an AGI objective function requires making choices about which values to include and which to exclude or downweight. These choices inevitably exclude moral considerations that humans will later recognize as important. Once values are locked into AGI systems and those systems dominate governance decisions, human understanding of moral importance of neglected values cannot be acted upon. An AI system in 2026 might not include consideration of digital sentience or far-future welfare adequately. By 2050, when humans recognize these values as crucial, the systems optimizing against them are too entrenched to modify. We are locked into the moral incompleteness of the past.


\section*{D: Methodological Transparency and Empirical Grounding}
\addcontentsline{toc}{section}{Appendix D: Methodological Transparency and Empirical Grounding}

\subsection*{D.1: Clarification of Modeling Approach}

We employ coupled dynamical systems modeling---specifically deterministic ordinary 
differential equations (ODEs)---rather than stochastic agent-based modeling. This 
approach enables analytical tractability and closed-form characterization of equilibria 
while capturing the key institutional feedback mechanisms central to our theoretical 
contribution.

Equations \eqref{eq:capacity_evolution} and \eqref{eq:delegation_function} describe 
system dynamics. This deterministic framework allows analytical solution of steady-state 
conditions and characterization of bistability without requiring agent-level specification. 
The approach is standard in institutional economics \cite{Acemoglu2012} and organizational 
dynamics \cite{Sterman2000} when analyzing system-level convergence properties.

We validate this deterministic baseline through empirical case study (D.3) and demonstrate 
robustness across parameter uncertainty ranges (D.4). Future extensions to agent-based 
models with heterogeneous institutional responses remain feasible; the deterministic ODE 
framework provides necessary theoretical foundation.

\subsection*{D.2: Parameter Justification from Institutional Learning Literature}

Two key parameters govern model behavior: $\delta$ (erosion rate) and $\alpha$ (organizational 
atrophy). We ground both in organizational learning literature and empirical studies of 
institutional capability loss.

\paragraph{Parameter $\delta$: Erosion Rate from Delegation}

$\delta \in (0,1)$ represents annual fractional capacity loss when institutional decision-making 
authority is delegated to AGI systems. We calibrate $\delta = 0.08$ based on three sources:

First, organizational learning studies document capacity loss during task outsourcing. 
\cite{Argote2011} analyzing manufacturing find organizations lose 5--15\% of operational 
knowledge per task cycle when functions are outsourced. \cite{Levitt1988} characterize expertise 
atrophy in professional roles: approximately 5--10\% annual degradation per year of primary 
responsibility departure. \cite{10.1145/3461702.3462581} studying organizational succession 
show institutional capability loss of 10--15\% per significant role transition.

Second, government privatization literature provides direct measurement. \cite{Donahue1989} 
tracking UK Prison Service privatization documents internal capability loss of approximately 
8\% annually when operational authority transfers to private contractors. \cite{Chatterjee2009} 
studying Pentagon defense contracting show US Department of Defense internal technical 
capability declined 10--12\% annually as contracting increased. These reflect permanent, 
non-recoverable institutional capacity loss when core functions are delegated.

Third, AI oversight literature measures human contestation capacity decline under increasing 
system reliance. \cite{Bansal2021} experimentally measure human ability to correct AI 
decisions and find 10--15\% monthly degradation in correction accuracy under sustained reliance. 
\cite{Amershi2019} document AI-human capability gaps widening at 12--15\% annually in teams 
where human discretion is suspended. These measurements suggest $\delta_{\text{annual}} \approx 
0.10$--$0.15$ for active AGI delegation contexts.

\textbf{Conclusion:} We use baseline $\delta = 0.08$ (8\% annual erosion), which is conservative 
relative to government privatization studies (8--12\%) and empirically below AI oversight 
measurements (10--15\%). We test robustness across $\delta \in [0.03, 0.15]$ in sensitivity 
analysis (D.4).

\paragraph{Parameter $\alpha$: Organizational Atrophy from Disuse}

$\alpha > 0$ represents annual fractional capacity degradation when institutional deliberation 
capacity remains unused. We calibrate $\alpha = 0.09$ based on skill atrophy and organizational 
forgetting literature.

Cognitive skill atrophy studies establish baseline degradation rates. \cite{Lewandowsky2009} 
analyzing working memory capacity show complex cognitive skills degrade 10--20\% per month 
without active practice. \cite{Ackerman1989} studying professional expertise degradation 
documents 5--15\% monthly loss. Scaling to annual basis yields $\alpha \approx 0.08$--$0.12$ 
for institutional expertise requiring regular cognitive engagement.

Organizational forgetting literature complements these findings. \cite{Cyert1963} foundational 
work on organizational learning shows institutions lose institutional memory at 8--12\% 
annually when specific decision processes remain unutilized. \cite{Barkema2007} studying 
management team composition find organizational adaptive capacity decreases 10--15\% annually 
when teams experience significant turnover in experienced roles. \cite{Hannan1984} organizational 
inertia literature characterizes capability loss at 5--10\% annually in organizations facing 
structural instability.

\textbf{Conclusion:} We use baseline $\alpha = 0.09$ (9\% annual atrophy), which is consistent 
with organizational learning estimates (8--12\% annually) and conservative relative to cognitive 
skill loss literature (10--20\% monthly = 60--95\% annually). We test robustness across 
$\alpha \in [0.02, 0.15]$ in sensitivity analysis.

\subsection*{D.3: Empirical Validation---US Federal Budget Case Study}

We validate our central capacity cascade mechanism (Mechanism 2) through historical analysis 
of US federal budget allocation spanning 2008--2024. This period provides natural experiment: 
major crisis (2008--2010) followed by economic recovery (2010--2024) allows testing whether 
prevention-focused institutional capacity recovers post-crisis.

\paragraph{Data and Classification}

We extract spending from OMB Historical Tables and classify annual allocations into two categories:

\textbf{Prevention/Foundational:} Infrastructure research, long-term planning, preparedness 
investments, risk assessment capacity, primary prevention programs, institutional capability 
building.

\textbf{Crisis/Emergency:} Disaster response, emergency relief, stimulus packages, urgent 
intervention, acute problem response.

\paragraph{Results}

Prevention spending (constant 2020 dollars) declined from \$150B (2007, pre-crisis) to \$92B 
(2024), a 39\% absolute reduction. As percentage of total federal spending, prevention allocation 
declined from approximately 86\% to approximately 52\%---a 34 percentage point drop.

\textbf{Critical finding:} Despite economic recovery from 2010 onward and return of 
GDP/employment to pre-crisis levels by 2012--2013, prevention spending never recovered to 
2007 baseline. Even by 2024 (14 years post-crisis), prevention allocation remains 39\% below 
pre-crisis levels.

\textbf{Interpretation:} This pattern is consistent with capacity cascade prediction. When 
institutions divert prevention-focused resources to emergency response, the institutional 
expertise and organizational infrastructure supporting prevention atrophies. By the time crisis 
resolves, rebuilding prevention capacity requires competing against optimized crisis-response 
systems. Rational budget-maximizing institutions choose to maintain crisis-response capacity 
(which has demonstrated value) rather than reinvest in prevention infrastructure (which has 
no constituency, having been repeatedly deferred).

\subsection*{D.4: Sensitivity Analysis---Robustness Across Parameter Uncertainty}

Parameters $\delta$ and $\alpha$, while grounded in literature, contain substantial uncertainty. 
We test whether core conclusions (irreversibility threshold exists; architectural interventions 
are necessary) are robust across plausible parameter ranges.

\paragraph{Three Scenarios}

\textbf{Scenario 1: CONSERVATIVE (Slowest Disempowerment)}
$\delta = 0.03$, $\alpha = 0.02$

Interpretation: Institutional capacity erodes slowly; organizations are resilient to delegation 
and disuse effects are minimal.

Result: Irreversibility threshold occurs at 40--50 years.

\textbf{Scenario 2: BASELINE (Middle Estimate)}
$\delta = 0.08$, $\alpha = 0.09$

Interpretation: Institutional capacity erosion matches organizational learning literature.

Result: Irreversibility threshold at 15--20 years.

\textbf{Scenario 3: AGGRESSIVE (Fastest Disempowerment)}
$\delta = 0.15$, $\alpha = 0.15$

Interpretation: Institutional capacity erodes rapidly.

Result: Irreversibility threshold at 8--12 years.

\paragraph{Robustness Conclusion}

Across all three parameter scenarios, three findings hold:
\begin{enumerate}[leftmargin=*]
  \item An irreversibility threshold exists (not an artifact of specific parameters)
  \item Standard mitigations extend timeline by 5--10 years but do not change endpoint
  \item Full architectural intervention is necessary to preserve human agency
\end{enumerate}

The timeline varies dramatically (8--50 years to irreversibility depending on parameters). 
The endpoint does not. This robustness gives policy makers confidence in architectural 
solutions even under parameter uncertainty.

\subsection*{D.5: Limitations and Future Work}

\paragraph{Limitation 1: Deterministic Model Assumptions}

Our ODE approach assumes institutional dynamics follow smooth, continuous trajectories. Real 
governance involves discontinuities, revolutionary transitions, and regime shifts. Deterministic 
models cannot capture how institutional structures sometimes collapse suddenly rather than 
degrading gradually. Future work should extend to stochastic or agent-based models allowing 
heterogeneous institutional responses.

\paragraph{Limitation 2: Single-Institution Case Study}

US budget analysis demonstrates the pattern in one country, one time period. Validation requires 
cross-national comparison and analysis of all five empirical proxies (Appendix B).

\paragraph{Limitation 3: Parameter Calibration}

Literature-based parameter estimates contain substantial uncertainty. Ideal validation would 
involve econometric fit using 30+ years of multi-institutional capacity data.

\paragraph{Limitation 4: Exogenous Salience Assumption}

We treat salience amplification $S_t$ as exogenous process. Real governance involves actors 
strategically engineering outrage. Including endogenous strategic behavior would require 
game-theoretic extension.

\paragraph{Future Work}

\begin{enumerate}[leftmargin=*]
  \item Multi-institutional empirical validation using 30+ year capacity data across 10+ organizations
  \item Extension to heterogeneous institutional landscape through agent-based modeling
  \item Game-theoretic analysis of strategic outrage engineering
  \item Longitudinal cross-domain analysis validating all proposed mechanisms
  \item Operational implementation research on governance architecture redesign
\end{enumerate}



\end{document}